\documentclass[12pt]{article}
\usepackage{amssymb,graphicx,epsfig,psfig}
\makeatletter\@addtoreset{equation}{section}\makeatother
\begin{document}

\begin{flushright}
{\tt hep-th/0601208}
\end{flushright}

\vspace{5mm}

\begin{center}
{{{\Large \bf Homogeneous Rolling Tachyons\\[2mm]
in\\[2mm]
Boundary String Field Theory}}\\[12mm]
{Akira Ishida, Yoonbai Kim, and Seyen Kouwn}\\[1mm]
{\it BK21 Physics Research Division and Institute of
Basic Science,\\
Sungkyunkwan University, Suwon 440-746, Korea}\\
{\tt ishida, yoonbai, seyen@skku.edu}
}
\end{center}
\vspace{10mm}

\begin{abstract}
We study decay of a flat unstable D$p$-brane in the context of
boundary string field theory action. Three types of homogeneous
rolling tachyons are obtained without and with Born-Infeld type
electromagnetic field.
\end{abstract}


\newpage

\setcounter{equation}{0}
\section{Introduction}

Rolling tachyons~\cite{Sen:2002nu} have been studied in the various contexts
including
boundary conformal field theory (BCFT), Dirac-Born-Infeld (DBI) type
effective field theory (EFT), noncommutative field theory (NCFT),
boundary string field theory (BSFT or background-independent string field
theory), cubic string field theory (SFT), and matrix models.
Various meaningful results were achieved in both homogeneous and
inhomogeneous rolling tachyon configurations and have been contributed
to understand real-time evolution of an unstable D-brane even in terms of
closed string degrees through the bridge of duality~\cite{Sen:2004nf}.

A cornerstone knowledge about the homogeneous rolling tachyons is on their
species, i.e., they are classified by three. In terms of BCFT, they are
hyperbolic cosine and sine type rolling tachyon
configurations~\cite{Sen:2002nu} called
as full S-branes~\cite{Gutperle:2002ai}, and exponential type rolling
tachyon~\cite{Larsen:2002wc} called as half S-brane~\cite{Strominger:2002pc}.
All of them share the same late-time behavior of vanishing
pressure~\cite{Sen:2002in}
so that classification is made by their different initial-time
behaviors.
This classification was also confirmed in DBI type EFT and NCFT even
in a form of exact solutions~\cite{Sen:2002an,Kim:2003he,Kim:2005pz}.
Though the late-time behavior of the rolling
tachyon represented by vanishing pressure is reproduced in
BSFT~\cite{Sugimoto:2002fp}, the simple observation on the three species of
rolling tachyons was not made in BSFT~\cite{Witten:1992qy} which
is appropriate to describe the open string tachyon
physics~\cite{Gerasimov:2000zp,Kutasov:2000aq}, a representative
off-shell dynamics in string theory. This may be related with complicated
form of the tachyon kinetic term in BSFT.

In this paper, we will find
all the three types of homogeneous rolling tachyons as classical solutions
of BSFT equations of motion. Comparison with other languages, particularly with
BCFT and EFT, shows complete agreement in spectrum of the homogeneous
rolling tachyons. The rolling tachyons with DBI type
electromagnetic coupling are also taken into account. We show that
the gauge field equation forces
its homogeneous field strength to be constant, and thus species of the
homogeneous rolling tachyons are the same as those without electromagnetic
coupling.

In section 2 we write down equations of motion from the BSFT tachyon action,
describing dynamics of an unstable D$p$-brane.
Section 3 devotes to the detailed analysis on homogeneous rolling tachyon
solutions for the case of the pure tachyon field, and reproduce all the three
types of homogeneous rolling tachyon solutions.
In section 4, we consider coupling of the DBI electromagnetic field which
stands for living of fundamental strings on the unstable D$p$-brane.
It is proved that every component of the homogeneous field strength
is constant for arbitrary $p$, and then inclusion of electromagnetism
does not change the spectrum composed of the three types of rolling tachyons.
We conclude in section 5 with brief discussions on further studies.

\setcounter{equation}{0}
\section{Tachyon Effective Action from BSFT}
We begin our discussion by recapitulating briefly derivation of an
effective action of a real tachyon
field~\cite{Kutasov:2000aq,Kraus:2000nj} and
abelian gauge field~\cite{Tseytlin:1987ww,Andreev:2000yn} in the
context of BSFT~\cite{Witten:1992qy},
and then read equations of motion.
In superstring theory, worldsheet partition function is believed to be
identified with off-shell BSFT action, $S=Z$~\cite{Marino:2001qc}.

For an unstable D$p$-brane, the partition function is given by
\begin{eqnarray}
Z=\int [{\cal D}X^{\mu}][{\cal D}\psi^{\mu}][{\cal D}{\bar \psi}^{\mu}]
\exp\left[-(I_{{\rm bulk}}+I_{{\rm bdry}})\right],
\end{eqnarray}
where $I_{{\rm bulk}}$ is the bulk action and $I_{{\rm bdry}}$ is the
boundary action of the form;
\begin{eqnarray}
I_{{\rm bdry}}= \int_{\partial\Sigma}\frac{d\tau}{2\pi}
\left[\frac{1}{4}T^{2}-\frac{1}{2}\psi^{\mu}\partial_{\mu}T
\partial_{\tau}^{-1}(\psi^{\nu}\partial_{\nu}T)
+\frac{i}{2} F_{\mu\nu}\psi^{\mu}\psi^{\nu}-\frac{i}{2}A_{\mu}
\frac{dX^{\mu}}{d\tau}
\right].
\end{eqnarray}
We consider a relevant linear perturbation for the tachyon
$T(X)= u_{\mu}X^{\mu}$ and constant electromagnetic field $F_{\mu\nu}$
with symmetric gauge $A_{\mu}=-F_{\mu\nu}X^{\nu}/2$ on the boundary.

Performing functional integration of the fields on the worldsheet and
using zeta function regularization, we reach
an action of the tachyon and the DBI electromagnetic field
\begin{eqnarray}\label{yac}
S(T,A_{\mu})= -{\cal T}_{p}\int d^{p+1}x \, V(T) \sqrt{-\det
(\eta_{\mu\nu}+F_{\mu\nu})}\, {\cal F}(y),
\end{eqnarray}
where the overall normalization was chosen by tension ${\cal T}_{p}$
of the unstable D$p$-brane, and
\begin{eqnarray}
V(T)&=&e^{-\frac{1}{4}T^{2}},\label{pot}\\
{\cal F}(y)&=&\frac{4^{y}y\, \Gamma(y)^{2}}{2\Gamma(2y)},\qquad
y=G^{\mu\nu}\partial_{\mu}T\partial_{\nu}T
\label{y}
\end{eqnarray}
with open string metric $G^{\mu\nu}=(\frac{1}{\eta+F})^{\mu\nu}_{\rm S}$.

From the action (\ref{yac}) we read the equations of motion for both the
tachyon and the gauge field. Then the obtained equations can be expressed
in terms of the open string metric $G^{\mu\nu}$ and noncommutativity parameter
$\theta^{\mu\nu}=(\frac{1}{\eta+F})^{\mu\nu}_{\rm A}$ as
\begin{eqnarray}\label{teq}
\partial_{\mu}\left[V\sqrt{-G}\, {\cal F'}
G^{\mu\nu}\partial_{\nu}T \right] =\frac{1}{2}\sqrt{-G}\, {\cal
F}\frac{dV}{dT},
\end{eqnarray}
\begin{eqnarray}\label{geq}
\partial_{\mu}\Pi^{\mu\nu}=0,
\end{eqnarray}
where
\begin{eqnarray}\label{pmn}
\Pi^{\mu\nu}&\equiv& \frac{\delta S}{\delta\partial_{\mu}A_{\nu}}
= {\tilde {\cal T}}_{p}V\sqrt{-G}\left[ \theta^{\mu\nu}{\cal
F} +2\left(\theta^{\rho\mu} G^{\sigma\nu} -\theta^{\rho\nu} G^{\sigma\mu}
\right){\cal F'}\partial_{\rho}T\partial_{\sigma}T\right].
\end{eqnarray}
Note that, in the previous expressions, the DBI Lagrangian with
string coupling $g_{{\rm s}}$ and slowly varying $F_{\mu\nu}$ in
(\ref{yac}) is equivalent to NCFT Lagrangian with open string
coupling $G_{{\rm s}}$ and nonconstant piece of noncommutative
field strength tensor ${\hat F}_{\mu\nu}$, up to total derivative
and ${\cal O}(\partial F)$ term~\cite{Seiberg:1999vs}
\begin{eqnarray}
-\frac{1}{g_{s}(2 \pi)^{\frac{p-1}{2}}} \sqrt{-{\rm
det}(\eta_{\mu\nu}+F_{\mu\nu})} = -\frac{1}{G_{s}(2
\pi)^{\frac{p-1}{2}}} \sqrt{-{\rm det}(G_{\mu\nu}+{\hat F}_{\mu\nu})} \,
\end{eqnarray}
with ${\rm det}(G_{\mu\nu}+{\hat F}_{\mu\nu})\equiv G$, ${\cal
T}_{ p}\equiv 1/g_{{\rm s}}(2\pi)^{\frac{p-1}{2}}$, and ${\tilde {\cal
T}}_{ p}\equiv 1/G_{{\rm s}}(2\pi)^{\frac{p-1}{2}}$.

Instead of the tachyon equation (\ref{teq}), it is economic to
examine conservation of energy-momentum tensor
\begin{equation}\label{emc}
\partial_{\mu}T^{\mu\nu}=0,
\end{equation}
where the energy-momentum tensor is also read from the action (\ref{yac})
\begin{eqnarray}\label{tmn}
\hspace{-5mm}T^{\mu\nu}&\equiv& \left.
\frac{2}{\sqrt{-g}}\frac{\delta S}{\delta
g_{\mu\nu}}\right|_{g_{\mu\nu}=\eta_{\mu\nu}}\nonumber\\
&=& -{\tilde {\cal
T}}_{p}V(T)\sqrt{-G} \biggr [ {\cal F}(y)
G^{\mu\nu}-2{\cal
F}'(y)(G^{\rho\mu}G^{\nu\sigma}+\theta^{\rho\mu}\theta^{\nu\sigma}
)\partial_{\rho}T\partial_{\sigma}T\biggr].
\end{eqnarray}

\setcounter{equation}{0}
\section{Homogeneous Rolling Tachyons}
In this section let us consider homogeneous configuration of the tachyon field
\begin{equation}\label{an}
T=T(t),
\end{equation}
without DBI electromagnetism, $F_{\mu\nu}=0$.
Then the equation of gauge field (\ref{geq}) becomes trivially satisfied
so that the only nontrivial equation is the tachyon equation (\ref{teq})
or equivalently the conservation of energy-momentum tensor (\ref{emc}).
Here we use the simpler conservation of energy-momentum tensor, which
reduces to $dT^{00}/dt=\dot{T}^{00}=0$ and it becomes
\begin{equation}\label{H}
{\cal H}\equiv T^{00}={\cal T}_{p}V(T)\left[{\cal F}(y)-2y{\cal F}'(y)\right],
\end{equation}
where $y=-\dot{T}^{2}\le 0$ from (\ref{y}).

\begin{figure}[ht]
\begin{center}
\scalebox{1}[1]{\includegraphics{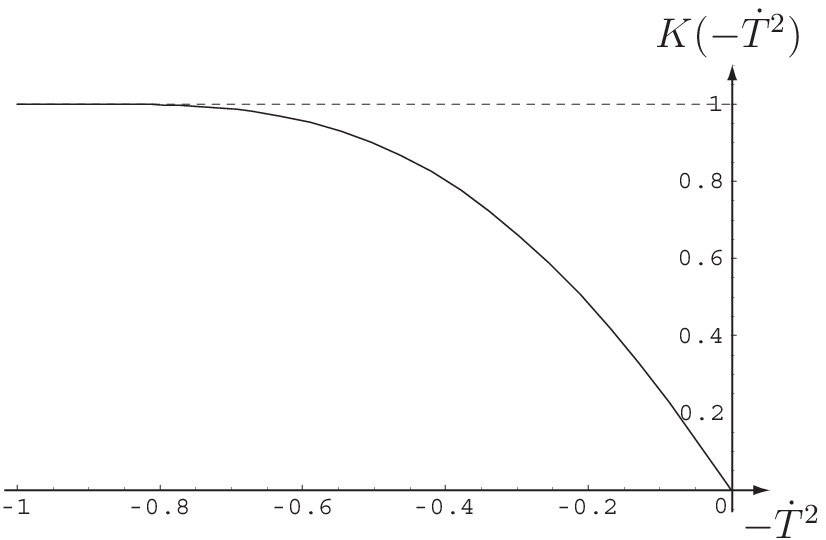}\includegraphics{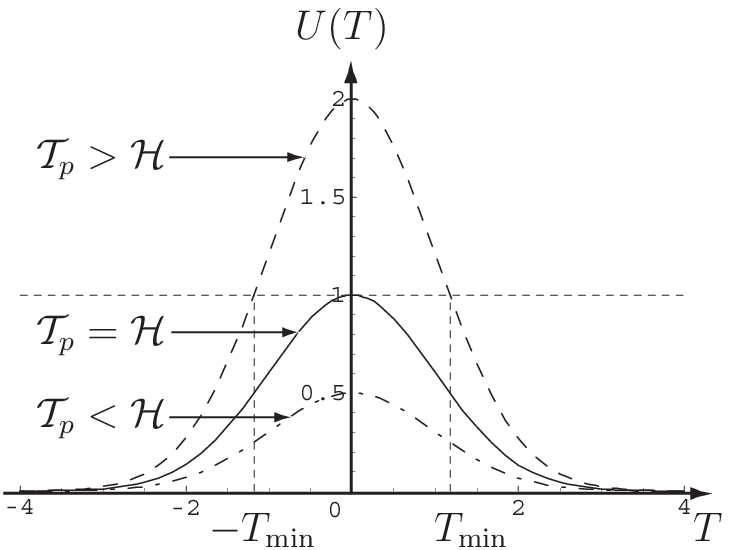}}
\par
\vskip-2.0cm{}
\end{center}
\caption{{\small The graphs of $K(y)$ (left) and $U(T)$ (right).
For $U(T)$, three representative cases of $({\cal T}_p / {\cal
H})^{2} =2~{\rm (dashed~curve)}, 1~{\rm (solid~curve)}, 1/2~\mbox{
(dotted-dashed~curve)}$ from top to bottom are shown. ${\cal E}=1$
is given by the straight dotted line.}}
\label{fig1}
\end{figure}

Reshuffling the terms of (\ref{H}), we obtain
\begin{equation}\label{eq}
{\cal E}=K(y)+U(T),
\end{equation}
where ${\cal E}=1$, and
\begin{eqnarray}
K(y)&=&1-\frac{1}{\left[{\cal F}-2y{\cal F}'\right]^{2}},
\label{K}\\
U(T)&=&\left(\frac{{\cal T}_{p}V}{{\cal H}}\right)^{2}
=\left(\frac{{\cal T}_{p}}{{\cal H}}\right)^{2}e^{-T^{2}/2}.
\label{U}
\end{eqnarray}
Though ${\cal F}(y)$ becomes divergent at each negative integer $y$,
the function $K(y)$ is bounded and behaves smoothly as shown in Fig.~\ref{fig1}.
Since we will use the range of $y$ from $-1$ to 0, $K(y)$ is monotonically
decreasing from max$(K(-1))=1$ to min$(K(0))=0$ in the $y$-range
of our interest. $U(T)$ is a runaway function having its maximum value
$({\cal T}_{p}/{\cal H})^{2}$ at $T=0$ and its minimum value ``zero''
at $T=\pm\infty$ as shown in Fig.~\ref{fig1}.
As given in Fig.~\ref{fig1}, homogeneous rolling tachyon solutions in BSFT
are classified by three cases. From now on we investigate detailed properties
of three types of rolling tachyons by examining the equation (\ref{eq}).
Since the energy density $T^{00}$ is a constant of motion and every
momentum density $T^{0i}$ and stress $T^{ij}~(i\ne j)$ vanishes,
physical property of each homogeneous rolling tachyon solution is expressed
by pressure component
\begin{eqnarray}
P&\equiv&T^{ii}\\
&=&-{\cal T}_{p}V(T){\cal F}(-{\dot T}^{2}).
\label{p}
\end{eqnarray}
To simplify our discussion without distorting physics, we use time
translation invariance so that the value of time $t$ has meaning
up to an overall constant like $t-t_{0}$.

\subsection{Hyperbolic cosine type rolling tachyon (${\cal T}_{p}>{\cal H}$)}

\begin{figure}[ht]
\begin{center}
\scalebox{1}[1]{\includegraphics{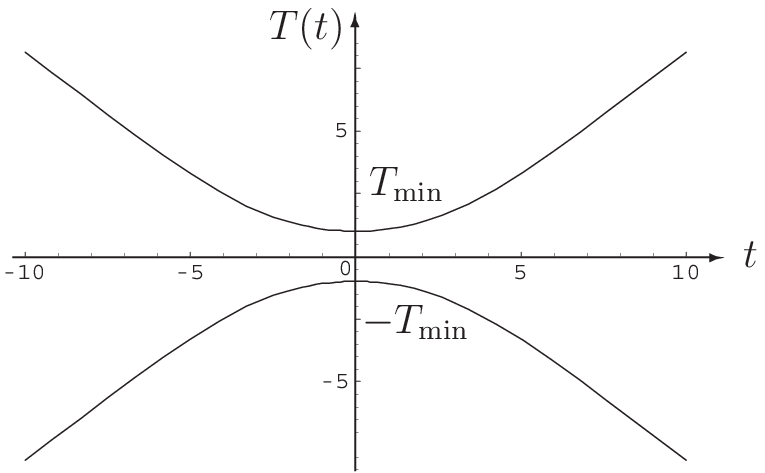}\includegraphics{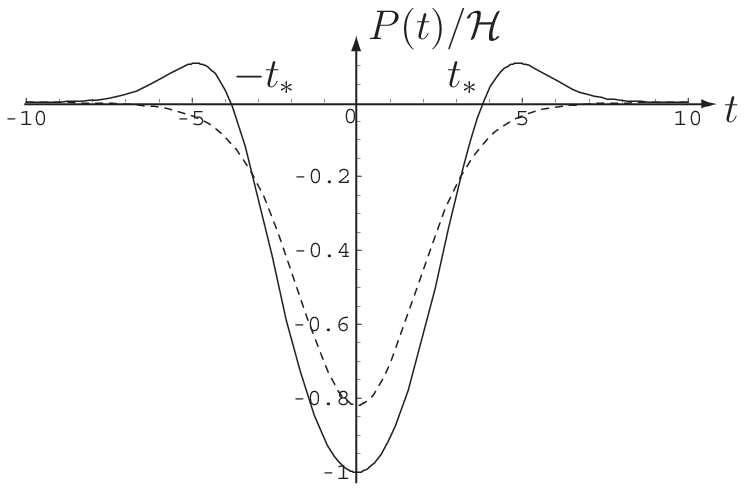}}
\par
\vskip-2.0cm{}
\end{center}
\caption{{\small The graphs of hyperbolic cosine function type
rolling tachyon (a full S-brane) for ${\cal T}_{p}>{\cal H}~({\cal
T}_{p}/{\cal H}=1.284)$: tachyon field $T(t)$ (left) and pressure
$P(t)/{\cal H}$ (right). The dashed line stands for the pressure
in BCFT.}} \label{cos}
\end{figure}

When ${\cal T}_{p}>{\cal H}$, it corresponds to the case of dashed curve in
Fig.~\ref{fig1}. The allowed range of the tachyon field is either
$T\ge T_{{\rm min}}=2\sqrt{\ln({\cal T}_{p}/{\cal H})}$ or
$T\le -T_{{\rm min}}$. Since the rolling tachyon of negative $T$ is signature
flipped copy of positive rolling tachyon, it is enough to describe the
positive case in what follows and the negative flipped copy will be given
only in figures without detailed explanation.
Possible homogeneous rolling tachyon is to start from
$T(-\infty)=\infty$, decreases monotonically to $T(0)=T_{{\rm min}}$
with ${\dot T}(0)=0$, and then return to $T(\infty)=\infty$. No analytic
solution is obtained but numerical solution is given in Fig.~\ref{cos}.
Shape of this rolling tachyon corresponds to the full S-brane of hyperbolic
cosine function.

Near the minimum value of the tachyon $T\approx T_{{\rm
min}}=2\sqrt{\ln({\cal T}_{p}/{\cal H})}$, $\dot T$ flips its
signature and has relatively small magnitude. and the kinetic term
(\ref{K}) is approximated as
\begin{equation}
K({-\dot T}^{2})\approx (4\log 2) \,\dot{T}^2 + {\cal O}(\dot{T}^4),
\end{equation}
Therefore, as the tachyon increases near $t=0$,
\begin{equation}\label{ct0}
T(t)\approx T_{{\rm min}}\left(1+\frac{t^2}{16\ln 2}+...\right),
\end{equation}
absolute value of the pressure (\ref{p}) decreases monotonically
\begin{eqnarray}\label{cp0}
\frac{P}{{\cal H}}\approx -1+\frac{T_{\rm min}^2}{16\ln 2} t^2+...
\,\,.
\end{eqnarray}
The leading unity comes from the ``cosmological constant" at the
turning point and the second positive term seems natural to reach
tachyon matter with vanishing pressure.

For sufficiently large $T\rightarrow \infty$, the tachyon
potential $V(T)$ and $U(T)$ decrease exponentially  zero, and
thereby the equation (\ref{eq}) with the help of the shape of
$K(y)$ in Fig.~\ref{fig1} dictates a universal behavior. Time
derivative of the tachyon arrives rapidly at unity
\begin{equation}\label{cti}
{\dot T}(t)\rightarrow 1-\frac{1}{2}\sqrt{\frac{{\cal
T}_{p}}{\cal{H}}}e^{-\frac{t^2}{8}},
\end{equation}
and, though ${\cal F}(y)$ diverges in this limit, character of the
pressure (\ref{p}) vanishes, dictated by the vanishing tachyon
potential;
\begin{eqnarray}\label{cpi}
P\sim \frac{\sqrt{{\cal T}_{p} \cal{H}}}{2}e^{-\frac{t^2}{8}}.
\end{eqnarray}
The late-time behavior is not a character of the hyperbolic cosine type
rolling
tachyon but a universal character of all the homogeneous rolling
tachyons in BSFT action~\cite{Sugimoto:2002fp},
shown by the solid lines in Figs.~\ref{cos}--\ref{sin}. Thus the
late-time behavior is different from that of pressure in BCFT,
which increases monotonically from
$P(t=0)=-{\cal H}/(2-{\cal H}/{\cal T}_{p})<0$ to
zero~\cite{Sen:2002nu,Larsen:2002wc} as shown by the dashed lines in
Figs.~\ref{cos}--\ref{sin}. This late-time behavior of the rolling
tachyon solutions, represented by vanishing pressure, is universal
irrespective of the detailed models and types of S-branes. This
so-called tachyon matter~\cite{Sen:2002in}, however it actually
reflects the matter behavior of vacuum of massive closed string
degrees~\cite{Lambert:2003zr}.

Since the pressure $P$ flips its signature as in (\ref{cp0}) and
(\ref{cpi}), there should exist a point of vanishing pressure
$P=0$ at a time $t=t_{\ast}$. From the expression of pressure
(\ref{p}), we read ${\cal F}(y=-{\dot T}_{\ast}^{2})=0$ and
thereby slope of the tachyon is fixed by ${\dot T}_{\ast}=\pm
\sqrt{1/2}\,$. Since ${\cal F}' (-1/2)=\pi$, value of the tachyon
field at the vanishing pressure is $T_{\ast}\equiv T(t_{\ast})=\pm
2\sqrt{\ln{({\cal T}_{p} \pi /{\cal H})}}$ from the equation
(\ref{H}). The pressure near $t=t_*$ is given by
\begin{equation}
 \frac{P}{{\cal H}}\approx \frac{1}{1+4\ln 2}\sqrt{\frac{1}{2}
 \ln\left(\frac{\pi {\cal T}_p}{\cal H}\right)}\,(t-t_*).
\end{equation}

\subsection{Exponential type rolling tachyon (${\cal T}_{p}={\cal H}$)}

\begin{figure}[ht]
\begin{center}
\scalebox{1}[1]{\includegraphics{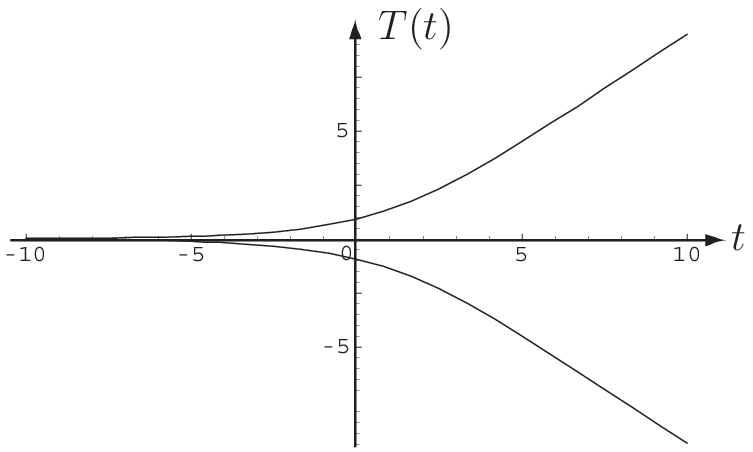}\includegraphics{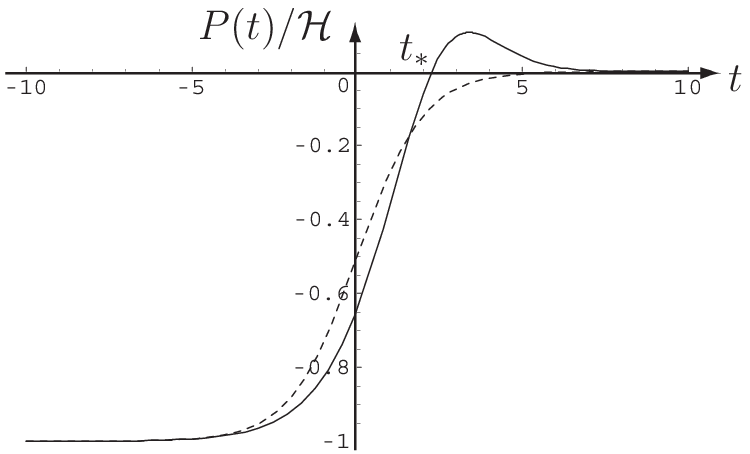}}
\par
\vskip-2.0cm{}
\end{center}
\caption{{\small The graphs of exponential function type rolling
tachyon (half S-brane) for ${\cal T}_{p}={\cal H}$: tachyon field
$T(t)$ (left) and pressure $P(t)/{\cal H}$ (right). The dashed
line stands for the pressure in BCFT.}} \label{exp}
\end{figure}

For the critical value of Hamiltonian density ${\cal H}={\cal
T}_{p}$, it corresponds to the solid curve in Fig.~\ref{fig1}.
Since $T_{{\rm min}} \rightarrow 0$, the rolling tachyon
configuration is a monotonically-increasing function from
$T(t=-\infty)=0$ to $T(t=\infty) =\infty$ and a solution through
numerical analysis is given in Fig.~\ref{exp}. Shape of the
obtained rolling tachyon is identified by exponential type
rolling tachyon in BCFT so-called half S-brane.

Even at initial stage near top of the tachyon potential at $T=0$
for $t\rightarrow -\infty$,
the tachyon field is already increasing exponentially like
the exact half S-brane configuration in BCFT
\begin{equation}\label{et0}
T(t) \approx \lambda e^{\frac{t}{2\sqrt{2\log{2}}}}+ ...,
\end{equation}
where $\lambda$ should be fixed by the boundary condition at
$t=\infty$. Insertion of it into the pressure $P$ (\ref{p})
provides leading negative contribution of the cosmological
constant at the top of tachyon potential and subleading tachyon
matter contribution
\begin{eqnarray}\label{ep0}
P\approx -{\cal T}_{p}\left(1-\frac{\lambda^2}{2}
e^{\frac{t}{\sqrt{2\log 2}}}+...\right).
\end{eqnarray}

At late time, leading behavior of this exponential type rolling
tachyon shares the same late-time behavior
(\ref{cti})--(\ref{cpi}) with the hyperbolic cosine type rolling
tachyon in the previous subsection (\ref{cti})--(\ref{cpi}) as
expected. The pressure $P(t)$ is also shown in Fig.~\ref{exp}.

\subsection{Hyperbolic sine type rolling tachyon (${\cal T}_{p}<{\cal H}$)}

\begin{figure}[ht]
\begin{center}
\scalebox{1}[1]{\includegraphics{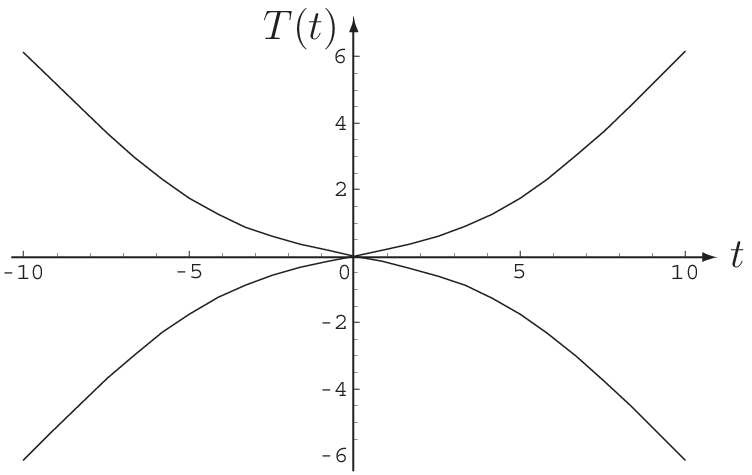}\includegraphics{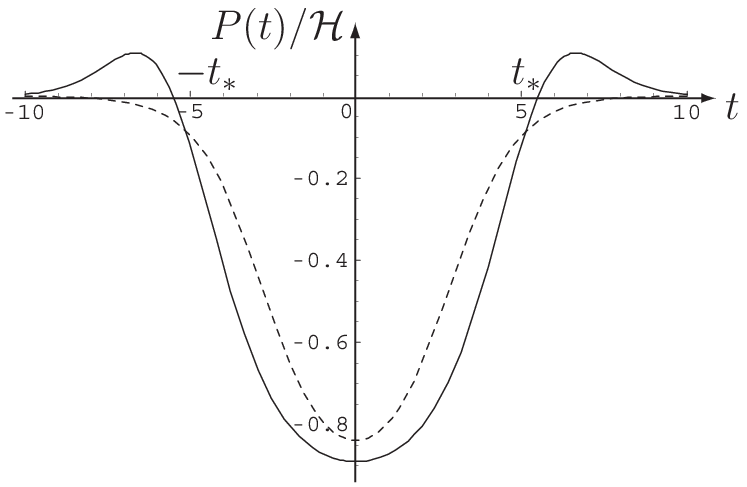}}
\par
\vskip-2.0cm{}
\end{center}
\caption{{\small The graphs of hyperbolic sine function type
rolling tachyon (full S-brane) for ${\cal T}_{p}<{\cal H}~({\cal
T}_{p}/{\cal H}=0.944)$: tachyon field $T(t)$ (left) and pressure
$P(t)/{\cal H}$ (right). The dashed line stands for the pressure
in BCFT.}} \label{sin}
\end{figure}

When ${\cal H}>{\cal T}_{p}$, ${\dot T}^2$ and the kinetic term
$K(-{\dot T}^2)$ are always positive so that the rolling tachyon
profile connects smoothly and monotonically one vacuum at
$T(t=-\infty)=\mp\infty$ and the other vacuum at
$T(t=\infty)=\pm\infty$, respectively, as shown by the numerical
solution in Fig.~\ref{sin}. It is identified as hyperbolic sine
type rolling tachyon corresponding to the other full S-brane.

Its late-time behavior is also universal and its initial-time
behavior at $t=-\infty$ is the reflected image of the late-time
behavior. When the tachyon field rolls over the top of tachyon
potential, expansion of $T(t)$ near $T(t=0)=0$ and
$\dot{T}(t=0)=\dot{T}_{0}$ provides
\begin{equation}\label{st0}
T(t) \approx \dot{T}_{0}\left( t + \frac{1}{24 C_0}
t^3\right)+{\mathcal O}(t^5),
\end{equation}
which is consistent with hyperbolic sine function near zero, and
\begin{eqnarray}\label{}
C_0 &\equiv& \frac{{\cal T}_{p}}{{\cal H}}\biggr[{\cal
F'}(-T_0^2)-2\dot{T}_0^2 {\cal F''}(-\dot{T}_0^2)\biggr]>0.
\end{eqnarray}
Then the pressure $P$ (\ref{p}) becomes
\begin{eqnarray}\label{sp0}
P\approx -{\cal T}_p \left[{\mathcal F}(-{\dot T}_0^2)
-\frac{{\dot T}_0^2}{4}\left({\mathcal F}(-{\dot
T}_0^2)-\frac{1}{C_0} {\mathcal F}'(-{\dot T}_0^2) \right)t^2
\right]+{\mathcal O}(t^4).
\end{eqnarray}

\vspace{5mm}

Among the homogeneous rolling tachyon configurations obtained in
the previous subsections, two of them in 3.1 and 3.3 are
identified with full S-branes~\cite{Gutperle:2002ai} and one in 3.2 is half
S-brane~\cite{Strominger:2002pc}. These completely coincide with
the results in BCFT~\cite{Sen:2002nu} and DBI type
EFT~\cite{Sen:2002an}. If all of these
theories are a variety of languages describing superstring theory
with an unstable D$p$-brane, there should exists a direct field
redefinition among the tachyon fields. In the subsequent section,
we analyze the cases with electromagnetic field turned on.

\setcounter{equation}{0}
\section{Homogeneous Rolling Tachyons with Electromagnetic Field}
When we derive the BSFT action (\ref{yac}), we assumed constant
electromagnetic field on the worldsheet.
On the unstable D$p$-brane
we take into account homogeneous configurations of the electromagnetic field,
consistent with the homogeneous tachyon field (\ref{an}),
\begin{equation}\label{anf}
F_{\mu\nu}=F_{\mu\nu}(t).
\end{equation}

In the subsequent subsections we analyze homogeneous rolling tachyon
solutions by examining the classical equations by
dividing the $p=1$ case and $p\ge 2$ cases,
since Bianchi identity
\begin{equation}\label{Bi}
\partial_{\mu}F_{\nu\rho}+\partial_{\nu}F_{\rho\mu}+\partial_{\rho}F_{\mu\nu}
=0,
\end{equation}
is trivial for $p=1$. The main results will be constancy of every component of
the electromagnetic field strength tensor, and then spectrum of the three
species of homogeneous rolling tachyon configurations will also be
kept to be the same as that of the pure tachyon case, irrespective of the
value of constant $F_{\mu\nu}$.

\subsection{$p=1$}
On an unstable D1-brane there exists only one electric component
of the field strength tensor $E(t)\equiv -F_{01}(t)$ and then we
have $-\det(\eta_{\mu\nu}+F_{\mu\nu})=1-E^{2}$. Because of the
homogeneity of the fields, (\ref{an}) and (\ref{anf}), momentum
density $T^{01}$ vanishes and then spatial component of the
energy-momentum conservation (\ref{emc}) is automatically
satisfied. The remaining time component of it leads to constancy
of the energy density (Hamiltonian density), $T_{00}={\cal
H}=$constant. Since $\Pi^{\mu\nu}$ (\ref{pmn}) is antisymmetric
due to antisymmetricity of $F_{\mu\nu}$, only the conjugate
momentum $\Pi\equiv \Pi^{01}$ survives. Thus, time component of
the gauge field equation (\ref{geq}) is trivially satisfied and
its spatial component results in constant conjugate momentum,
$\Pi=$constant. Specific expressions of the Hamiltonian ${\cal
H}=T_{00}$ (\ref{tmn}) and the conjugate momentum $\Pi=\Pi^{01}$
tell us that ratio of two quantity is negative electric field $E$,
\begin{equation}\label{Ee}
{\cal H}=-\frac{\Pi}{E}=\frac{{\cal T}_{p}V}{\sqrt{1-E^{2}}}
\left[{\cal F}-2y{\cal F}'\right],\qquad y=-\frac{{\dot
T}^{2}}{1-E^{2}}
\end{equation}
which forces constant electric field.

When $1-E^{2}>0$, rescaling of the time as
$\tilde{t}=t\sqrt{1-E^{2}}$ and the Hamiltonian density ${\tilde
{\cal H}}=\sqrt{1-E^{2}}\,{\cal H}$ lets the equation (\ref{Ee})
the same equation (\ref{H}) of pure tachyon case. Since the
rescaling preserves the boundary conditions, this proves that
spectrum of the homogeneous rolling tachyon solutions with
homogeneous electric field coincides exactly with that without
electromagnetic field in the previous section 3 through the
rescaling of time ${\tilde t}$ and Hamiltonian density ${\tilde
{\cal H}}$.

Vanishing electric field limit $E^{2}=0$ corresponds obviously to pure
tachyon limit.
For critical electric field $E^{2}=1$, the rescaling becomes singular
and then we need careful analysis. If we rewrite the equation (\ref{Ee}),
then we have
\begin{equation}\label{1}
{\cal E}_{1}=K_{1}(y)+U_{1}(T),
\end{equation}
where
\begin{equation}\label{2}
{\cal E}_{1}=1,\qquad
K_{1}(y)=1-\frac{1}{\left[{\cal
F}(y)-2y{\cal F}'(y)\right]^{2}},\qquad
U_{1}=\frac{{\cal T}_{p}^{2}V^{2}}{(1-E^{2}){\cal H}^{2}}.
\end{equation}
In the critical limit $E^{2}\rightarrow 1$, $U_{1}$ becomes divergent unless
$T\rightarrow \pm \infty$ with $V(T=\pm \infty)=0$. From the variable $y$
in (\ref{Ee}), $y\rightarrow -\infty$ unless ${\dot T}\rightarrow 0$.
Thus there exists a trivial vacuum solution of $T=\pm \infty$ and ${\dot T}=0$
in the critical limit of the electric field. If ${\dot T}\ne 0$, $y\rightarrow
-\infty$ and then
${\displaystyle \lim_{y\rightarrow -\infty}K_{1}(y)\rightarrow 1}$.
The equation (\ref{1})
forces $V(T)=0$ and subsequently the Hamiltonian in (\ref{Ee}) can be finite.
This nontrivial rolling vacuum solution attained due to the vacuum at
infinity is not overruled.
Correspondingly, for both the trivial vacuum solution and the rolling
vacuum solution, the pressure becomes constant as
\begin{eqnarray}\label{Ep}
\lim_{E^{2}\rightarrow 1}P=\lim_{E^{2}\rightarrow 1}
-\frac{{\cal T}_{p}V}{\sqrt{1-E^{2}}}\left[
{\cal F}-2E^{2}y{\cal F}'\right]= -{\cal H}.
\end{eqnarray}

\subsection{$p\ge 2$}
The homogeneous electric field $E$ on the unstable D1-brane is proven to be
constant due to classical equations of motion, and a rescaling of the time
variable led to the same spectrum of homogeneous rolling tachyon solutions,
i.e., they are hyperbolic cosine and sine type rolling tachyons
(two full S-branes), and exponential type rolling tachyon (a half S-brane).
For higher-dimensional D$p$-brane, number of components of the field
strength tensor increases as $p(p+1)/2$ and the number of equations of
motion with Bianchi identity (\ref{Bi}) also increase.
In this subsection, through a careful analysis, all the components of
homogeneous electromagnetic fields are proven to be constant
and, through an appropriate rescaling of time variable, the single tachyon
equation supports exactly the same three homogeneous rolling tachyon
solutions.

We begin the analysis with homogeneity of the fields (\ref{an})
and (\ref{anf}) for the cases of $p\ge 2$. For the homogeneous
fields, the Bianchi identity (\ref{Bi}) requires every component
of the magnetic field $F_{ij}$ to be constant, i.e., $p(p-1)/2$
components among $p(p+1)/2$ components are constants. In addition,
the conservation of the energy-momentum (\ref{emc}) becomes
simple, ${\dot T}^{0\nu}=0$, which lets $T^{0\nu}$ constants
\begin{equation}\label{p1}
T^{0\nu}=\frac{{\cal T}_{p}V}{\sqrt{\beta}}C^{0\nu}_{{\rm S}}
\left[{\cal F}(y)-2y{\cal F}'(y)\right]
={\rm constant},
\end{equation}
where $y$ is, from (\ref{y}),
\begin{equation}
y=-\alpha{\dot T}^{2}/\beta .
\end{equation}
Throughout this subsection we use notations that
$\beta=-\det(\eta_{\mu\nu}+F_{\mu\nu})$, $C^{\mu\nu}$ denotes
cofactor of the matrix $(\eta+F)_{\mu\nu}$ with its symmetric part
$C^{\mu\nu}_{{\rm S}}$ and antisymmetric part $C^{\mu\nu}_{{\rm
A}}$, and $\alpha=C^{00}$. Since $C^{00}_{{\rm S}}$ contains only
the magnetic field $F_{ij}$ which is constant, it is also a
constant. Combining this with (\ref{p1}), we obtain
\begin{equation}\label{p2}
\frac{{\cal T}_{p}V}{\sqrt{\beta}}\left[{\cal F}(y)-2y{\cal
F}'(y)\right]=\frac{T^{00}}{C^{00}}=\frac{{\cal H}}{\alpha},
\end{equation}
where the Hamiltonian density ${\cal H}$ is a constant. Similarly,
the equation of the gauge field (\ref{geq}) makes every charge
density of the fundamental strings constant, i.e., ${\dot
\Pi}^{0\nu}=0$ results in constant $\Pi^{0i}$
\begin{equation}\label{p3}
\Pi^{0i}=\frac{{\cal T}_{p}V}{\sqrt{\beta}}C^{0i}_{{\rm A}}
\left[{\cal F}(y)-2y{\cal F}'(y)\right]=\frac{{\cal
H}}{\alpha}C^{0i}_{{\rm A}}
\end{equation}
which results in constant $C^{0i}_{{\rm A}}$. Therefore, as far as the
determinant $\beta$ is nonvanishing, the $p$ independent relations,
$\left(\frac{1}{\eta+F}\right)^{0i}_{{\rm A}}=C^{0i}_{{\rm A}}/\beta$,
require every electric field
$E^{i}=-F_{0i}$ is also constant. Summarized the results of $p=1$
case in the subsection 4.1 and of $p\ge 2$ cases in the
subsection 4.2, we now show that all the field strength
components assumed to be homogeneous are constants on every
unstable D$p$-brane.

Now that the only remaining first-order equation (\ref{p2}) is equivalent
to the tachyon
equation (\ref{teq}), we rewrite it in a convenient form like (\ref{eq})
\begin{equation}\label{p4}
{\cal E}_{p}=K_{p}(y)+U_{p}(T),
\end{equation}
where
\begin{equation}\label{p5}
{\cal E}_{p}=1,\qquad
K_{p}(y)=1-\frac{1}{\left[{\cal
F}(y)-2y{\cal F}'(y)\right]^{2}},\qquad
U_{p}=\frac{{\alpha^2\cal T}_{p}^{2}V^{2}}{\beta{\cal H}^{2}}.
\end{equation}
Since $\alpha=C^{00}$ is positive and
$\beta=-\det(\eta_{\mu\nu}+F_{\mu\nu})$ is nonnegative, the
equation (\ref{p4}) supports the same three types of homogeneous
rolling tachyon solutions. To be specific, they are hyperbolic
cosine type rolling tachyon solution for ${\sqrt{\beta}\,{\cal
H}}/{\alpha}<{\cal T}_{p}$, exponential type rolling tachyon
solution for ${\sqrt{\beta}\,{\cal H}}/{\alpha}={\cal T}_{p}$, and
hyperbolic sine type rolling tachyon solution for
${\sqrt{\beta}\,{\cal H}}/{\alpha}>{\cal T}_{p}$. The obtained
spectrum of rolling tachyons in BSFT action coincides with that in
BCFT~\cite{Mukhopadhyay:2002en},
DBI type EFT~\cite{Mukhopadhyay:2002en,Gibbons:2002tv,Kim:2003he},
and NCFT~\cite{Mukhopadhyay:2002en,Kim:2005pz}. Even functional
form of the tachyon field for the homogeneous rolling tachyons is
the same up to the rescaling of the time variable so that we omit
detailed analysis of their physical properties.

\setcounter{equation}{0}
\section{Conclusion}

We studied homogeneous rolling tachyons in the framework of
the BSFT action and classified the obtained solutions into three cases,
i.e., they are hyperbolic cosine, exponential, and hyperbolic sine type
configurations. Analytic solutions were not obtained in closed form,
but numerical solutions coincide with our analysis (refer to the
Figs.~\ref{cos}-\ref{sin}). The late-time shapes of the pressure
commonly approach to zero for sufficiently late time.

We also considered homogeneous rolling tachyon solutions
in the presence of homogeneous electromagnetic field.
Bianchi identity and the equation of motion for the gauge field
require every component of the electromagnetic field to be constant.
Then the nonzero electromagnetic field turns out to lead to the rescaling
of the time variable. Therefore all of the homogeneous rolling
solutions to the BSFT equations of motion are classified into three
cases irrespective of existence of the DBI electromagnetic field.

Now the discussions for further studies are in order.
A noteworthy difference between BCFT results and the obtained
BSFT results is on behavior of the pressure.
There exists a finite time where the pressure vanishes
and sign of it flips in BSFT, however the pressure in BCFT
keeps its negative signature for all the time.
In relation with this discrepancy,
it will be intriguing to consider the rolling tachyons by
taking into account the correction from higher derivative terms.

Since all the perturbative open string modes on an unstable D-brane disappear
in the late-time of its decay~\cite{Ishida:2002fr},
production of nonperturbative
soliton solutions through inhomogeneous rolling
tachyons~\cite{Ishida:2003cj} is worth tackling.

\section*{Acknowledgements}
We would like to thank Chanju Kim, O-Kab Kwon, and Ho-Ung Yee for helpful
discussions.
This work is the result of research activities (Astrophysical
Research Center for the Structure and Evolution of the Cosmos
(ARCSEC)) supported by Korea Science $\&$ Engineering Foundation
and was supported by Samsung Research Fund, Sungkyunkwan
University, 2005. The work of A.I. is also supported in part by
the Postdoctoral Research Program of Sungkyunkwan University
(2005).

\end{document}